# Encryption of Binary and Non-Binary Data Using Chained Hadamard Transforms

Rohith Singi Reddy
Computer Science Department
Oklahoma State University, Stillwater, OK 74078

**ABSTRACT**

This paper presents a new chaining technique for the use of Hadamard transforms for encryption of both binary and non-binary data. The lengths of the input and output sequence need not be identical. The method may be used also for hashing.

**INTRODUCTION**

The Hadamard transform is an example of generalized class of Fourier transforms [4], and it may be defined either recursively, or by using binary representation. Its symmetric form lends itself to applications ranging across many technical fields such as data encryption [1], signal processing [2], data compression algorithms [3], randomness measures [6], and so on.

Here we consider an application of the Hadamard matrix to non-binary numbers. To do so, Hadamard transform is generalized such that all the values in the matrix are non negative. Each negative number is replaced with corresponding modulo number; for example while performing modulo 7 operations -1 is replaced with 6 to make the matrix non-binary. Only prime modulo operations are performed because non prime numbers can be divisible with numbers other than 1 and itself. The use of modulo operation along with generalized Hadamard matrix limits the output range and decreases computation complexity.

In the method proposed in this paper, the input sequence is split into certain group of bits such that each group bit count is a prime number. As we are dealing with non-binary numbers, for each group of binary bits their corresponding decimal value is calculated. The input sequence is split in such a way that, the maximum decimal value thus obtained by the group is a prime number. The reason to consider only prime number is because our operations are on modulo prime numbers after converting into the non-binary form.

Both during encryption and decryption the sequence after converting into the non-binary form is broken down into chunks of equal size as a power of 2. Moreover, the whole input sequence is divided into groups of equal size and each group is then multiplied with the corresponding Hadamard matrix. The Hadamard matrix which is to be multiplied is chosen based on the maximum value of the decimal sequence. For example in a sequence of groping of three bits, the maximum value that can be obtained is $2^3-1=7$, and hence Hadamard matrix of modulo 7 is used to perform the operations corresponding to the sequence.

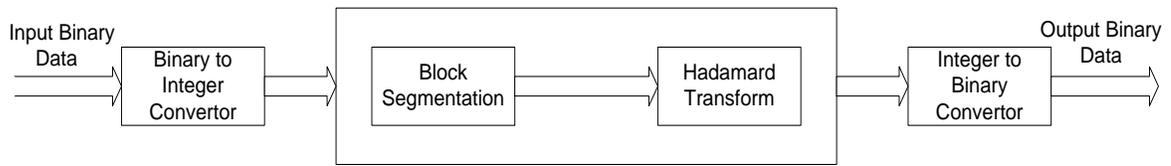

Figure 1. The Encryption System

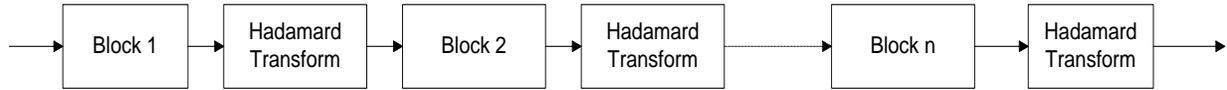

Figure 2. Chaining in Block Sequence which expands the middle block of Figure1

The general scheme of the proposed encryption system is shown in Figure 1 and Figure 2. The power of this is a consequence of the many different ways and the number of times the chaining of blocks is done. Similar pattern can be generated for decryption in the reverse order.

Initially before performing any operations, all the numbers in the decimal sequence need to be analyzed. As, modulo number of the matrix is equal to the maximum number obtained by grouping the sequence, there is a possibility for the sequence to have the maximum values, for example 111 produce 7 which is equal to modulo number of the matrix. To overcome this problem a two dimensional integer array is used. Before performing any operations, at each level the array is filled with 0s or 1s in such a way that, for each maximum value in the decimal sequence 1 is entered in the array corresponding to the index of the maximum value and remaining other indexes are filled with 0s. This scheme of representation helps to keep track of all the maximum numbers in the sequence, if any. And during decryption whenever 0 is encountered in the sequence the corresponding position is checked in the array, if the position shows 1 in the array then 0 in the sequence is replaced with maximum value for that particular sequence.

**ENCRYPTION**

The idea of encryption is to multiply the decimated input sequence with the non-binary Hadamard matrix in a chained manner. The encryption is carried out at number of levels which makes it hard to break the encrypted message. Here, the number of levels in which the encryption is carried will depend on the key sequence provided, for example key sequence with numbers {3, 5, 7} will perform encryption at three levels one for each number in the key. Moreover, the given key sequence should contain only prime numbers such that $2^x-1$ is also a prime where x is any number in the sequence. This sequence can be considered as the public key based on which both encryption and decryption are carried.

Initially, the first number from the sequence is considered and whole input sequence is grouped based on the number. The decimal values for each group obtained thereafter are calculated. As we are dealing with Hadamard matrix which is a square matrix, the decimated input sequence is divided into chunks of equal size corresponding to the size of the matrix. Once after dividing the decimal sequence and if the size of the sequence is less than size of the matrix then zeros must be appended at the end of the decimal sequence to make it equal to the matrix size. Appending zeros is done at each level for those sequences whose length is less than the matrix size.

After getting the decimal sequence for the first number in the key, the decimal sequence obtained is then represented as a column matrix. Next, the Hadamard matrix is multiplied with the decimal sequence column matrix. The matrix which is to be multiplied by the input sequence depends on the number in the given key. For example, if the number in key is 3 then the matrix representing modulo 7 must be used to perform the multiplication. Therefore, at each level different combinations of the Hadamard matrix is used to perform encryption. Since, the multiplication can generate large values, so to limit the output and to make sure that the original sequence is obtained after decryption, modulo operation is applied on the resultant matrix.

The resultant matrix obtained by performing modulo operation is then converted to the binary sequence and this specifies the end of Level 1. Each level uses the binary sequence obtained from its previous level and follows the same encryption procedure. This process is continued until all the numbers in the key are processed and the binary sequence thus obtained will be the final encrypted message. The final encrypted message may have a length which is equal or different from the input sequence.

**Encryption Algorithm**

Step1: Firstly, consider the given binary input sequence and the key. The given key can have n numbers such that each number say x in n is a prime and $2^x-1$ is also a prime. Also, consider a two dimensional integer array. Here, number of rows is equal to number of elements in the array and number of columns is equal to number of input values at each row.

Step2: Next, consider the first element in the key and group the bits in the input sequence based on the number.

Step3: Now convert each group to corresponding decimal number.

Step4: Depending on the length of input sequence, divide the decimated sequence to equal length sub-sequences such that each sub-sequence length should be expressed as power of 2.

Step5: If length of the sub-sequence cannot be expressed as power of 2, append $0^s$ at last to make the length equal to other sub sequences length.

Step6: For every sub-sequence, if a number in the sub-sequence is equal to $2^x-1$, then corresponding index in the array is marked with 1 and remaining all indexes are marked with $0^s$.

Step7: Represent each sub-sequence as a column matrix. Multiply each sub-sequence matrix with the modified Hadamard matrix, such that the matrix of the form modulo $2^x-1$ must be used to perform multiplication.

Step 8: Apply modulo $2^x-1$ on the resultant values obtained after multiplication.

Step 9: Convert each decimal number in the sequence to corresponding binary values.

Step 10: Now, consider next element in the key and group the bits of the sequence obtained from the previous step based on the number.

Step 11: Repeat Step4 to Step9. The sequence obtained after processing the last element in the key is the final encrypted message.

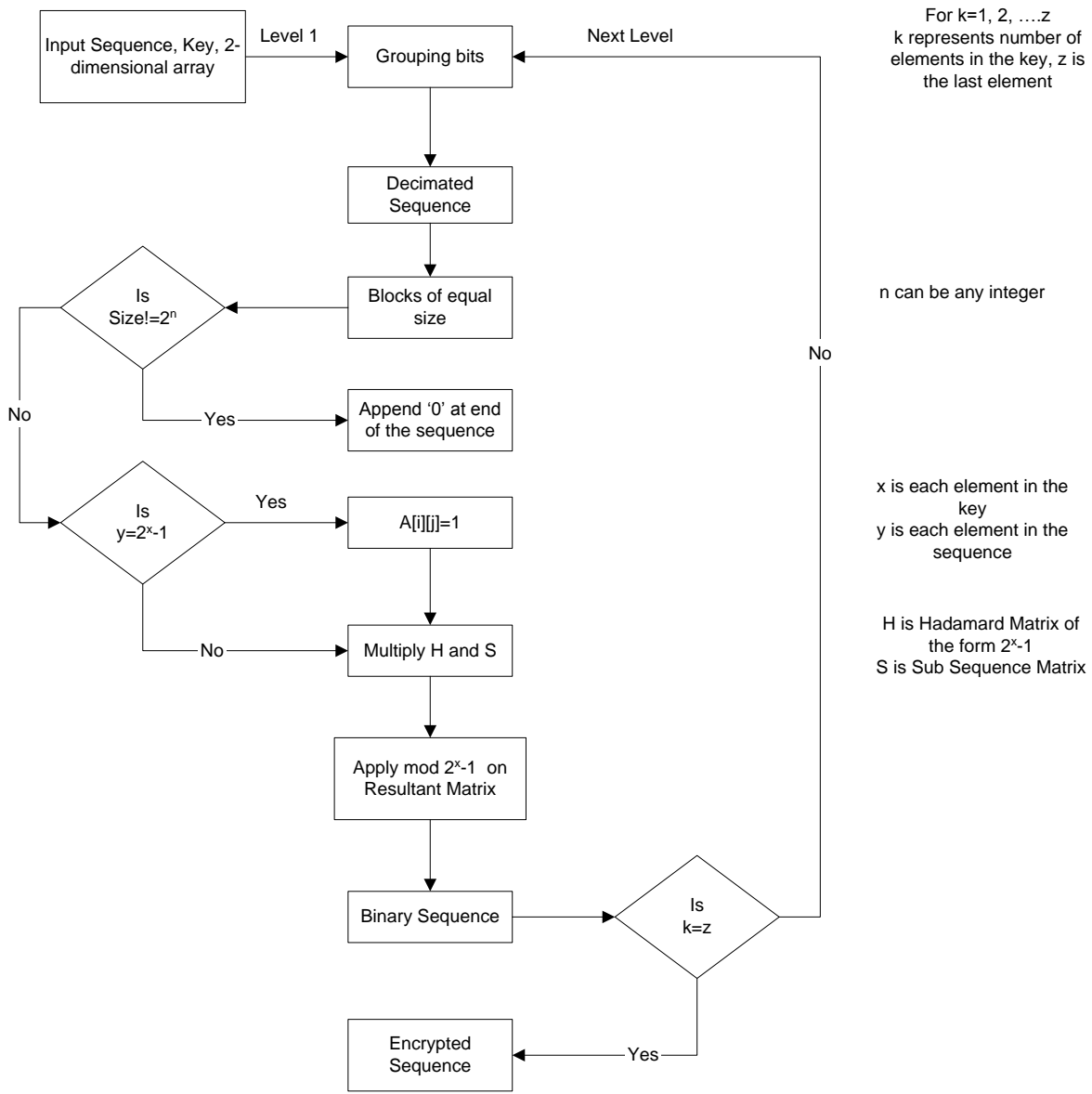

Figure 3. Block Diagram for Encryption

**Note:** In case if any bits are padded at Level 1 the count for the bits must be noted and after finishing the final level in decryption the last bits of the sequence must be discarded based on the count noted in Level 1 of encryption.

**DECRYPTION**

The decryption scheme discussed in this paper uses the same Hadamard matrix as in encryption with some additional operations. In order to get back the original sequence the decryption should be performed in a reversed order. Hence, the key sequence is reversed and decryption is performed at levels on each element in the key, for example key of sequence {3, 5, 7} is reversed as {7, 5, 3} and similar to encryption each element in the key is considered and decryption is performed. The decryption method is started by considering the first element in the key and whole encrypted sequence is then grouped based on the number selected. Now, decimated values for each group is calculated and later the sequence is divided into equal sized sub-sequences as done in encryption. Once after division is done, each sub-sequence is then represented by column matrix. In order to perform multiplication, the Hadamard matrix corresponding to modulo $2^x-1$ is chosen, for example if grouping is done on element 3 then Hadamard matrix corresponding to $2^3-1$ i.e, modulo 7 must be chosen.

After getting the decimated sequence each column matrix is multiplied with the Hadamard matrix, and to get the original message the column matrix must be multiplied with the inverse Hadamard matrix. The inverse Hadamard matrix is same as the original Hadamard matrix along with an additional multiplier operation. To perform decryption using the next element in the key, the decimated output is represented in the binary form.

During encryption we append extra 0s at end of the sequence, so these 0s should be discarded to get the original sequence. Here, only those 0s at the end of sequence which are consecutive must be discarded and make sure that the modified sequence thus obtained should be represented as power of 2. While performing encryption all the maximum values in the sequence are stored in the array, so here in decryption at each level for every occurrence of 0 in the sequence, corresponding index is checked in the array and if the index has a value 1 then 0 in the sequence is replaced by the maximum value for that sequence.

Each level thereafter uses the output obtained from the previous level and performs similar operations on it. Thus, the binary sequence obtained after processing the final element in the key will be the original message.

**Decryption Algorithm**
Step1: Initially, consider the encrypted sequence, reversed key sequence and the array used in encryption.

Step2: Next, consider the first element in the key and group the bits in the encrypted sequence based on the number.

Step3: Now convert each group to corresponding decimal number.

Step4: Depending on the length of input sequence, divide the decimated sequence to equal length sub-sequences such that each sub-sequence length should be expressed as power of 2.

Step5: Represent each sub-sequence as a column matrix. Now, multiply each sub-sequence matrix with the modified Hadamard matrix, such that the matrix of the form modulo $2^x$-1 must be used to perform multiplication.

Step6: Now, multiply two modulo $2^x$-1 matrices and find the divisor such that the resultant matrix obtained is thus represented as an Identity matrix.

Step7: Calculate modulo multiplicative inverse for the divisor that is, a*y mod $2^x$-1 =1[5] where y is the divisor and a is modulo multiplicative inverse.

Step8: Multiply the resultant matrix obtained in Step7 with a.

Step9: Apply modulo $2^x$-1 on the resultant values obtained after multiplication.

Step10: For every 0 in the decimal sequence check for the corresponding array index, if the index has an element 1 then replace 0 with $2^x$-1.

Step11: Convert each decimal number in the sequence to corresponding binary values.

Step12: Remove all consecutive 0s at end of the sequence such that, the resultant sequence has a length equal to power of 2.

Step13: Now, consider next element in the key and group the bits of the sequence obtained from the previous step based on the number.

Step14: Repeat Step4 to Step12. The sequence obtained after processing the last element in the key is the original message.

## SIMULATION

For simplicity only modulo 7 and modulo 31 operations are considered during encryption and decryption. So, the key can have combinations of numbers 3 and 5. Initially, the combinations of these matrices ie, 8×8, 16×16, 32×32 are stored in an array and corresponding matrix is chosen depending on input length and key value.

**Example 1**
Hadamard matrix corresponding to modulo 7

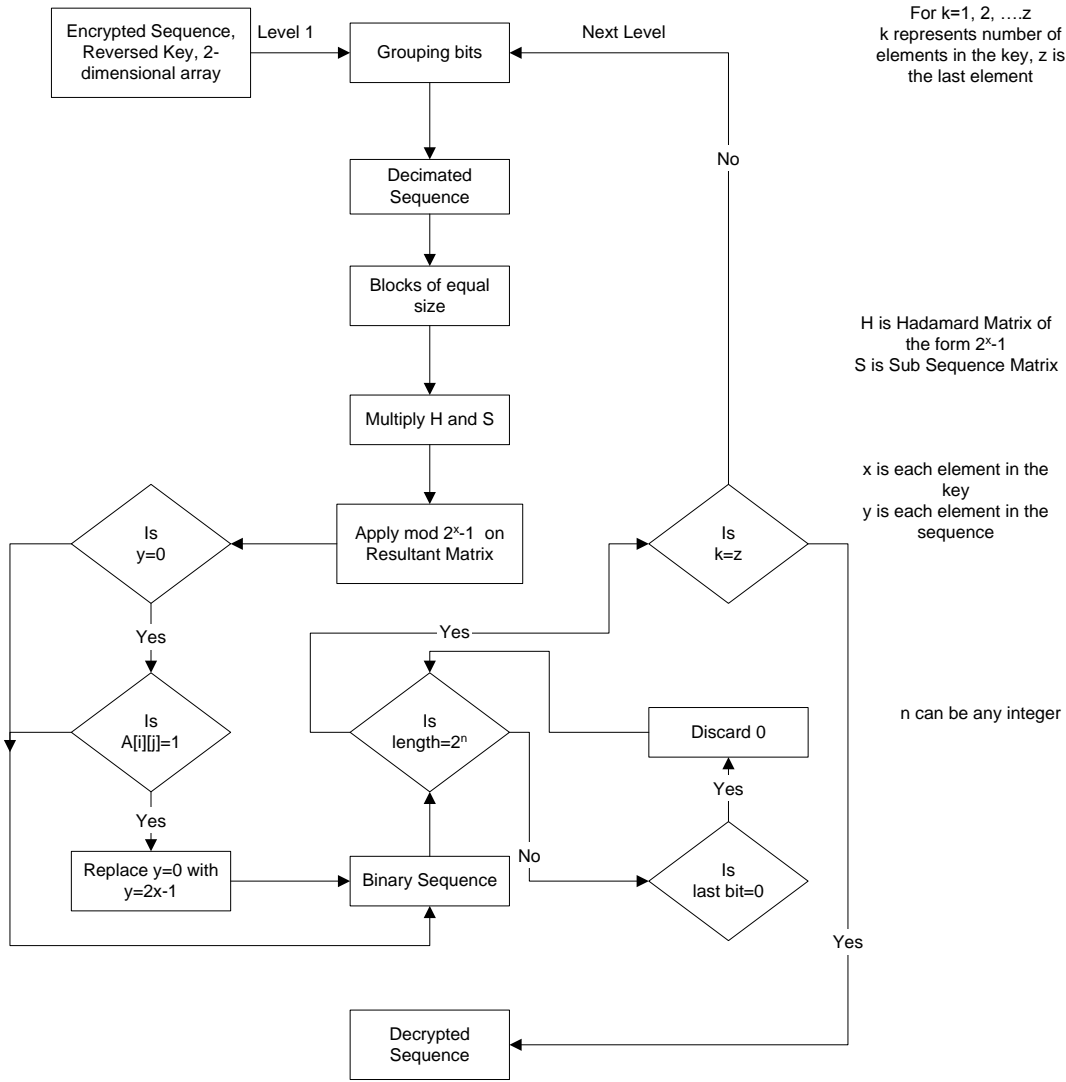

Figure 4. Block Diagram for Decryption

$$\begin{pmatrix} 1 & 1 & 1 & 1 & 1 & 1 & 1 & 1 \\ 1 & 6 & 1 & 6 & 1 & 6 & 1 & 6 \\ 1 & 1 & 6 & 6 & 1 & 1 & 6 & 6 \\ 1 & 6 & 6 & 1 & 1 & 6 & 6 & 1 \\ 1 & 1 & 1 & 1 & 6 & 6 & 6 & 6 \\ 1 & 6 & 1 & 6 & 6 & 1 & 6 & 1 \\ 1 & 1 & 6 & 6 & 6 & 6 & 1 & 1 \\ 1 & 6 & 6 & 1 & 6 & 1 & 1 & 6 \end{pmatrix}$$

Hadamard matrix corresponding to modulo 31

$$\begin{pmatrix} 1 & 1 & 1 & 1 & 1 & 1 & 1 & 1 \\ 1 & 30 & 1 & 30 & 1 & 30 & 1 & 30 \\ 1 & 1 & 30 & 30 & 1 & 1 & 30 & 30 \\ 1 & 30 & 30 & 1 & 1 & 30 & 30 & 1 \\ 1 & 1 & 1 & 1 & 30 & 30 & 30 & 30 \\ 1 & 30 & 1 & 30 & 30 & 1 & 30 & 1 \\ 1 & 1 & 30 & 30 & 30 & 30 & 1 & 1 \\ 1 & 30 & 30 & 1 & 30 & 1 & 1 & 30 \end{pmatrix}$$

**Encryption Part**
Input Binary Sequence: 110010011101111110000011
Key {3, 5}

**Level1**
At Level 1 every 3 bits in the input sequence is grouped and corresponding decimal value is calculated.

Decimated Sequence: 6, 2, 3, 5, 7, 6, 0, 3
The maximum value corresponding to the input sequence is stored in the array.
A[0][0]={0, 0, 0, 0, 1, 0, 0, 0}

Multiplying Hadamard matrix and decimated sequence and then performing the modulo operation, we get the following sequence
4, 0, 3, 3, 0, 4, 4, 4, 2

Converting to Binary: 100000011011000100100010

**Level 2**
Modified Binary Sequence: 10000001101100010010001000000000000000
At Level 2 every 5 bits in the sequence obtained from above level is grouped and corresponding decimal value is calculated.

Decimated Sequence: 16, 6, 24, 18, 4, 0, 0, 0

Multiplying Hadamard matrix and decimated sequence and then performing the modulo operation, we get the following sequence
6, 20, 15, 8, 29, 12, 7, 0

Converting to Binary: 00110101000111101000111010110000111 00000
Encrypted Message: 00110101000111101000111010110000111 00000

Figure 8 given below shows difference between the original sequence and the encrypted sequence. By looking at the graph it can be visualized that the encrypted message obtained thereafter is completely different from the original input sequence.

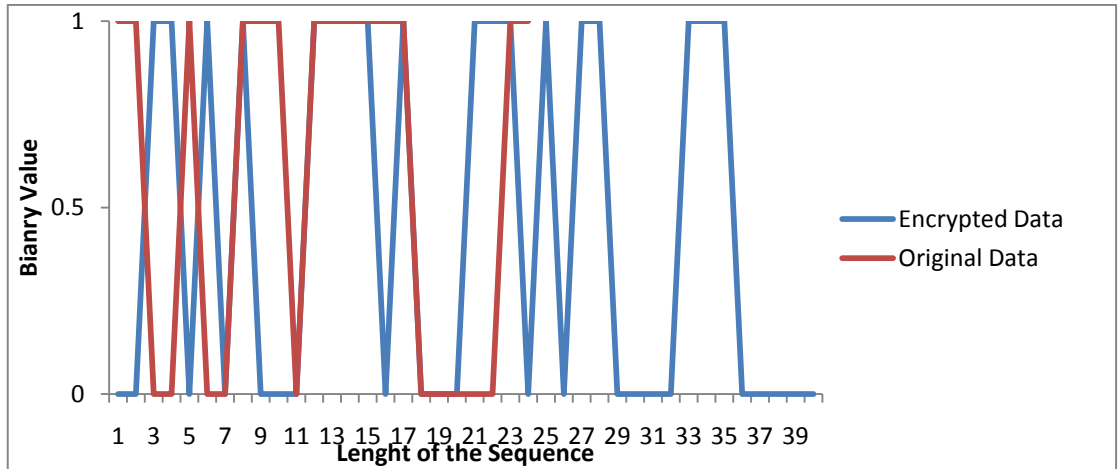
Figure 8. Original Sequence Vs Encrypted Sequence (Example 1)

**Decryption Part**
Key= {5, 3}

**Level1**
At Level 1 every 5 bits in the encrypted sequence is grouped and corresponding decimal value is calculated.
Input Sequence: 0011010100011110100011101011000011100000
Decimated Sequence: 6, 20, 15, 8, 29, 12, 7, 0

Multiplying Hadamard matrix and decimated sequence, we get
97, 1257, 967, 1663, 1489, 1953, 1953, 1953
Calculating Modulo Multiplicative Inverse

$$\begin{pmatrix} 1 & 1 & 1 & 1 & 1 & 1 & 1 & 1 \\ 1 & 30 & 1 & 30 & 1 & 30 & 1 & 30 \\ 1 & 1 & 30 & 30 & 1 & 1 & 30 & 30 \\ 1 & 30 & 30 & 1 & 1 & 30 & 30 & 1 \\ 1 & 1 & 1 & 1 & 30 & 30 & 30 & 30 \\ 1 & 30 & 1 & 30 & 30 & 1 & 30 & 1 \\ 1 & 1 & 30 & 30 & 30 & 30 & 1 & 1 \\ 1 & 30 & 30 & 1 & 30 & 1 & 1 & 30 \end{pmatrix} \times \begin{pmatrix} 1 & 1 & 1 & 1 & 1 & 1 & 1 & 1 \\ 1 & 30 & 1 & 30 & 1 & 30 & 1 & 30 \\ 1 & 1 & 30 & 30 & 1 & 1 & 30 & 30 \\ 1 & 30 & 30 & 1 & 1 & 30 & 30 & 1 \\ 1 & 1 & 1 & 1 & 30 & 30 & 30 & 30 \\ 1 & 30 & 1 & 30 & 30 & 1 & 30 & 1 \\ 1 & 1 & 30 & 30 & 30 & 30 & 1 & 1 \\ 1 & 30 & 30 & 1 & 30 & 1 & 1 & 30 \end{pmatrix} =$$

$$\begin{pmatrix} 8 & 124 & 124 & 124 & 124 & 124 & 124 & 124 \\ 124 & 3604 & 1922 & 1922 & 1922 & 1922 & 1922 & 1922 \\ 124 & 1922 & 3604 & 1922 & 1922 & 1922 & 1922 & 1922 \\ 124 & 1922 & 1922 & 3604 & 1922 & 1922 & 1922 & 1922 \\ 124 & 1922 & 1922 & 1922 & 3604 & 1922 & 1922 & 1922 \\ 124 & 1922 & 1922 & 1922 & 1922 & 3604 & 1922 & 1922 \\ 124 & 1922 & 1922 & 1922 & 1922 & 1922 & 3604 & 1922 \\ 124 & 1922 & 1922 & 1922 & 1922 & 1922 & 1922 & 3604 \end{pmatrix}$$

$$1/8 \times \begin{pmatrix} 8 & 124 & 124 & 124 & 124 & 124 & 124 & 124 \\ 124 & 3604 & 1922 & 1922 & 1922 & 1922 & 1922 & 1922 \\ 124 & 1922 & 3604 & 1922 & 1922 & 1922 & 1922 & 1922 \\ 124 & 1922 & 1922 & 3604 & 1922 & 1922 & 1922 & 1922 \\ 124 & 1922 & 1922 & 1922 & 3604 & 1922 & 1922 & 1922 \\ 124 & 1922 & 1922 & 1922 & 1922 & 3604 & 1922 & 1922 \\ 124 & 1922 & 1922 & 1922 & 1922 & 1922 & 3604 & 1922 \\ 124 & 1922 & 1922 & 1922 & 1922 & 1922 & 1922 & 3604 \end{pmatrix} \mod 31 =$$

$$\begin{pmatrix} 1 & 0 & 0 & 0 & 0 & 0 & 0 & 0 \\ 0 & 1 & 0 & 0 & 0 & 0 & 0 & 0 \\ 0 & 0 & 1 & 0 & 0 & 0 & 0 & 0 \\ 0 & 0 & 0 & 1 & 0 & 0 & 0 & 0 \\ 0 & 0 & 0 & 0 & 1 & 0 & 0 & 0 \\ 0 & 0 & 0 & 0 & 0 & 1 & 0 & 0 \\ 0 & 0 & 0 & 0 & 0 & 0 & 1 & 0 \\ 0 & 0 & 0 & 0 & 0 & 0 & 0 & 1 \end{pmatrix}$$

=> a*8 mod 31=1
=> Modulo multiplicative inverse is 4
Matrix after multiplying with modulo multiplicative inverse is
388, 5028, 3868, 6652, 5956, 7812, 7812, 7812

Sequence after performing modulo 31
16, 6, 24, 18, 4, 0, 0, 0
All the 0 values are checked against the array and as there is no maximum value stored, so the sequence is not modified.
16, 6, 24, 18, 4, 0, 0, 0

Converting to Binary: 10000001101100010010000000000000000000

**Level2**
At Level 2 every 5 bits in the sequence obtained from above level is grouped and corresponding decimal value is calculated.
Discarding Additional 0s: 1000000110110001001000010
Decimated Sequence: 4, 0, 3, 3, 0, 4, 4, 2

Multiplying Hadamard matrix and decimated sequence, we get
20, 65, 80, 75, 70, 55, 70, 45

Calculating Modulo Multiplicative Inverse

$$\begin{pmatrix} 1 & 1 & 1 & 1 & 1 & 1 & 1 & 1 \\ 1 & 6 & 1 & 6 & 1 & 6 & 1 & 6 \\ 1 & 1 & 6 & 6 & 1 & 1 & 6 & 6 \\ 1 & 6 & 6 & 1 & 1 & 6 & 6 & 1 \\ 1 & 1 & 1 & 1 & 6 & 6 & 6 & 6 \\ 1 & 6 & 1 & 6 & 6 & 1 & 6 & 1 \\ 1 & 1 & 6 & 6 & 6 & 6 & 1 & 1 \\ 1 & 6 & 6 & 1 & 6 & 1 & 1 & 6 \end{pmatrix} \times \begin{pmatrix} 1 & 1 & 1 & 1 & 1 & 1 & 1 & 1 \\ 1 & 6 & 1 & 6 & 1 & 6 & 1 & 6 \\ 1 & 1 & 6 & 6 & 1 & 1 & 6 & 6 \\ 1 & 6 & 6 & 1 & 1 & 6 & 6 & 1 \\ 1 & 1 & 1 & 1 & 6 & 6 & 6 & 6 \\ 1 & 6 & 1 & 6 & 6 & 1 & 6 & 1 \\ 1 & 1 & 6 & 6 & 6 & 6 & 1 & 1 \\ 1 & 6 & 6 & 1 & 6 & 1 & 1 & 6 \end{pmatrix} =$$

$$\begin{pmatrix} 8 & 28 & 28 & 28 & 28 & 28 & 28 & 28 \\ 28 & 148 & 98 & 98 & 98 & 98 & 98 & 98 \\ 28 & 98 & 148 & 98 & 98 & 98 & 98 & 98 \\ 28 & 98 & 98 & 148 & 98 & 98 & 98 & 98 \\ 28 & 98 & 98 & 98 & 148 & 98 & 98 & 98 \\ 28 & 98 & 98 & 98 & 98 & 148 & 98 & 98 \\ 28 & 98 & 98 & 98 & 98 & 98 & 148 & 98 \\ 28 & 98 & 98 & 98 & 98 & 98 & 98 & 148 \end{pmatrix}$$

$$1 \times \begin{pmatrix} 8 & 28 & 28 & 28 & 28 & 28 & 28 & 28 \\ 28 & 148 & 98 & 98 & 98 & 98 & 98 & 98 \\ 28 & 98 & 148 & 98 & 98 & 98 & 98 & 98 \\ 28 & 98 & 98 & 148 & 98 & 98 & 98 & 98 \\ 28 & 98 & 98 & 98 & 148 & 98 & 98 & 98 \\ 28 & 98 & 98 & 98 & 98 & 148 & 98 & 98 \\ 28 & 98 & 98 & 98 & 98 & 98 & 148 & 98 \\ 28 & 98 & 98 & 98 & 98 & 98 & 98 & 148 \end{pmatrix} \mod 7$$

$$= \begin{pmatrix} 1 & 0 & 0 & 0 & 0 & 0 & 0 & 0 \\ 0 & 1 & 0 & 0 & 0 & 0 & 0 & 0 \\ 0 & 0 & 1 & 0 & 0 & 0 & 0 & 0 \\ 0 & 0 & 0 & 1 & 0 & 0 & 0 & 0 \\ 0 & 0 & 0 & 0 & 1 & 0 & 0 & 0 \\ 0 & 0 & 0 & 0 & 0 & 1 & 0 & 0 \\ 0 & 0 & 0 & 0 & 0 & 0 & 1 & 0 \\ 0 & 0 & 0 & 0 & 0 & 0 & 0 & 1 \end{pmatrix}$$

=> a*1 mod 7=1
=> Modulo multiplicative inverse is 1
Matrix after multiplying with modulo multiplicative inverse
20, 65, 80, 75, 70, 55, 70, 45

Sequence after performing modulo 7
6, 2, 3, 5, 0, 6, 0, 3
All the 0 values are checked against the array and 0 at fifth position is replaced with 7.
Modified Sequence after replacing with maximum value is
6, 2, 3, 5, 7, 6, 0, 3

Converting to Binary: 110010011101111110000011

Decrypted Message: 110010011101111110000011

**Example 2**
**Changing one bit in the decrypted message**

Input Sequence: 110010011101111110000011
Encrypted Sequence: 001101010001111010001110101100001110000
Changed Encrypted Sequence: 101101010001111010001110101100001110000
Decrypted Sequence: 010010010001111000100000010000011101000000000110

In the example discussed above, the starting bit in the encrypted sequence is changed from 0 to 1and a drastic difference is observed between the original sequence and the decrypted sequence. Moreover, not only the bits in the decrypted part are changed but also there is varied length between the sequences.

If the input sequence contains all 1s or 0s then the corresponding encrypted sequence is observed to contain all 0s which implies the input does not carry any useful information.

In the proposed scheme structure of the key plays a vital role. It will be challenging for the eavesdropper to break the sequence if the length of the key is large enough and contain distinct elements. On the other hand if the key length is short or if it has many duplicate elements then it can be easily broken using brute force or other techniques.

**CONCLUSION**

This paper presents an approach to encryption using a chain of Hadamard transforms. We have presented implementation of this system for binary and non-binary message sequence. An algorithm is given for both encryption and decryption along with the block diagram. Clearly, this method is more effective when the key size is large and most of the elements in the key are distinct. But since the chaining of the Hadamard transforms can be as long as one pleases and since the choices as far as the formation of blocks that are converted into non-binary numbers is arbitrarily large, the effectiveness of this scheme is potentially very high.

The technique of chaining can be applied using other transforms also. The proposed method may be used for hashing by padding the given data block with zeros, finding the encryption sequence, and retaining a certain pre-specified number of bits. Such a procedure can then serve as a method of joint encryption and error-correction coding [7].

**REFERENCES**

1. L. J. Yan and J. S. Pan, Generalized discrete fractional Hadamard fransformation and its application on the image encryption, International Conference on Intelligent Information Hiding and Multimedia Signal Processing, IEEE Computer Press, pp. 457-460, 2007.